\documentclass[prd,showpacs,amsmath,amssymb,preprintnumbers,
10pt,superscriptaddress]{revtex4}

\usepackage{graphicx}
\usepackage{dcolumn}
\usepackage{bm}
\usepackage{epsfig} 
\begin{document}


\title{Supergravity can reconcile dark matter with lepton number
violating neutrino masses}

\author{Biswarup Mukhopadhyaya}
\email{biswarup@mri.ernet.in}
\affiliation{Harish-Chandra Research Institute, Chhatnag Road, Jhusi,
Allahabad - 211 019, India}
\author{Soumitra SenGupta}
\email{tpssg@iacs.res.in}
\affiliation{Department of Theoretical Physics, Indian Association for
the Cultivation of Science,
Kolkata - 700 032, India}
\author{Raghavendra Srikanth}
\email{srikanth@mri.ernet.in}
\affiliation{Harish-Chandra Research Institute, Chhatnag Road, Jhusi,
Allahabad - 211 019, India}

\begin{abstract}
Supersymmetry offers a cold dark matter candidate, provided that
lepton number is {\em not violated by an odd number of units}. 
On the other hand, lepton number violation by even (two) units gives
us an attractive
mechanism of neutrino mass generation. Here we offer an explanation
of this, in a supergravity framework
underlying a supersymmetric scenario, the essential feature being
particles carrying lepton numbers, which interact only gravitationally 
with all other known particles. It is shown that one can have the
right amount of $\Delta L = 2$ effect giving rise to neutrino
masses, whereas the lifetime for $\Delta L = 1$ decays of the
lightest supersymmetric particle can be prolonged beyond the
present age of the universe.


\end{abstract}


\maketitle

\section{Introduction}

Speculations abound nowadays about a supersymmetric (SUSY) nature
at the fundamental level, 
described by a theory invariant under boson-fermion transformations.  
One of the positive features of SUSY \cite{susy} is that in
its minimal form it provides a candidate for cold dark matter in our
universe, since, in most models, the
lightest SUSY particle (LSP) is stable, neutral and
weakly interacting only, and also lies in the mass range
of the electroweak scale \cite{dm}. However, the stability of
the LSP  requires the SUSY theory to conserve R-parity, defined as
$R = (-1)^{L+3B+2s}$, where $L$, $B$ and $s$ stand for the lepton number, 
baryon number and spin of a particle, respectively. Arguing in this line,
lepton number is expected to be conserved in order to ensure that SUSY 
is the source of cold dark matter.

To be very precise, however, the conservation of R-parity means that
$L$ is not violated  {\em by an odd number of units}. Thus one may 
conserve R-parity and retain stability of the LSP even if 
$L$ is violated by {\em even number of units}, which makes 
it possible to have $\Delta L = 2$ neutrino masses of the form
${\bar\nu^c} \nu$. Such mass terms form the seed for, say, the
seesaw mechanism \cite{seesaw} which is a beautiful explanation of the
smallness of neutrino masses vis-a-vis the 
masses of the charged leptons. So the smallness of neutrino
masses and the dark matter in the universe can be explained
together in SUSY models if we assume lepton number is violated
by two units. In contrast to this, with $\Delta L = 1$ terms one
may explain the smallness of neutrino masses but additional
sources for dark matter are required to be postulated. The
question that one may ask now is: if  $\Delta L = 2$ is allowed, is there
any fundamental reason to believe that $\Delta L = 1$ terms
either cannot occur or are very suppressed?
In other words, can a cold dark matter candidate be reconciled with
Majorana masses for neutrinos with the help of some fundamental principles?
These questions form the central theme of the present work.

The superpotential of a lepton number conserving theory, including
right-handed neutrino superfields $N_i$ (required for 
the seesaw mechanism), is
\begin{equation}
W_{MSSM} = Y_u^{ij}Q_iU^c_jH_2 + Y_d^{ij}Q_iD^c_jH_1 
	  + Y_e^{ij}L_iE^c_jH_1 + 
	Y_{\nu}^{ij}L_iN^c_jH_2 - \mu H_1H_2,
	\label{eq:mssm}
\end{equation}
where the flavor indices, $i,j$ run from 1 to 3 and SU(2) gauge
indices have been suppressed. The $Y$'s stand for various Yukawa couplings.
$\mu$ is the Higgsino mass parameter.
$H_1,H_2$ are the two SU(2) doublet Higgs superfields, with
$H_1 = (H_1^0,H_1^-)$ and $H_2 = (H_2^+,H_2^0)$.
$Q,L$ are SU(2) doublet quarks and leptons, while 
$U^c,D^c,E^c$ are SU(2) singlet up-quark, down-quark
and charged lepton superfields respectively.
If $L$ is violated, then one can further add the terms \cite{rpv}
\begin{equation}
W_{\not{L}} = \lambda_{ijk}L_iL_jE^c_k + \lambda^\prime_{ijk}Q_iL_jD^c_k
	      + \epsilon_iL_iH_2 + M_{ij}N^c_iN^c_j,
	\label{Lnot}
\end{equation}
where $\lambda,\lambda^\prime$ are some constants and $\epsilon,M$ are
mass parameters. Here the first three terms violate $L$ by one unit, 
and need to be forbidden
for the stability of LSP. The last term, violating $L$ by two units,
gives Majorana masses for neutrinos. The aim, therefore, is to try to
understand why the last term should be allowed but not the first three
terms of equation (\ref{Lnot}).

There have been some explanations of the above claim in, say, supersymmetric
Grand Unified Theories (GUTs). For example, in a SUSY GUT model based on
the gauge group $SU(5)\times SU(5)$, R-parity can automatically be
conserved and seesaw masses for neutrinos can also be generated \cite{gut}.
There have been some SUSY models where R-parity arises
from a continuous gauge symmetry \cite{Martin}. Since R-parity is
naturally conserved, the stability of LSP and Majorana nature of neutrinos
can be understood in these models.

Here we take an alternative approach and seek an explanation of such a
phenomenon in a supergravity (SUGRA) framework which is by far the most
popular paradigm of SUSY breaking \cite{sugra}. Apart from being
the local extension of global SUSY, SUGRA can also have its root in
some radically fundamental idea like superstring theory. In this framework, 
the SUSY-breaking soft terms have their origin in non-renormalizable 
interactions of the
observable fields with a hidden sector which is sterile under all known
interactions excepting gravity. Also, non-renormalizable
interactions with the hidden sector have often been invoked to explain
neutrino masses \cite{ahmsw,by,ms}.
We utilize this framework to explain the Majorana nature of neutrinos
and the dark matter content of the universe. As described below, we
postulate some symmetries applicable to hidden sector fields, suppresses
phenomenological effects of the
unwanted $\Delta L = 1$ terms and explain why lepton number is violated by
two units in the neutrino sector. Unlike the models \cite{gut,Martin},
where R-parity is naturally conserved, this model leads to R-parity
violating terms when the global symmetries (lepton number and R-charge)
are spontaneously broken below the Planck scale. We show that the effect
of these R-parity violating terms is small and
the lightest neutralino of MSSM can be a long lived particle. The lifetime
of this neutralino field can be larger than the age of the universe and
can serve as a candidate for dark matter.
The two central features used by us are
(a) the possibility of some hidden sector field carrying  lepton number, and
(b) the fact that the theory may have a
non-minimal Kahler potential, thus accommodating certain desirable values
of parameters in the observable sector.

In the proposed scenario, we make use of three hidden sector chiral superfields
$S (L = 0)$, $S^\prime (L = 0)$ and $X (L = 1)$, where $L$ within the
brackets indicate the lepton number of the field. As it will be shown
in detail later, the purpose of the fields $S,S^\prime$ is to give masses
to the scalar and gaugino fields. Among the hidden sector fields, $X$
carries lepton number +1, and we propose this field to establish the
Majorana nature of neutrinos. While it may be unusual
to attribute lepton number to a hidden sector field, such suggestions have
been considered earlier in the context of leptogenesis \cite{bb}. It should
also be remembered that a right-handed neutrino superfield itself is a
gauge singlet carrying lepton number, and can pass off as a hidden sector
field but for the Yukawa interactions. However, the lepton number assignments 
prevent $X$ from entering into Yukawa couplings. To generate $\Delta L = 2$
terms for neutrinos, we assume that lepton number is conserved at the
Planck scale. This assumption forbids the usual $N^cN^c$ term in the
superpotential, but a non-renormalizable term $\frac{XX}{M_P}N^cN^c$ is
allowed. If $X$ acquires a vacuum expectation value (vev) $\Delta L = 2$
terms can be generated for neutrinos, and to have right-handed neutrino
mass at the TeV scale the vev of $X$ should be at the intermediate scale
($10^{10-11}$ GeV). The assumption of lepton number conservation may
not be enough to postulate, since terms like $XN^c$ are also allowed in the
superpotential, and they generate unwanted $\Delta L = 1$ terms
through the vev of $X$. To avoid these problems, we make use of the
R-symmetry, assumed to be conserved at the Planck scale. The breakdown
of these symmetries at somewhat lower scales is motivated by (a) the
requirement of seesaw masses for neutrinos and (b) the need to achieve
adequate suppression of $\Delta L = 1$ effects when the symmetries
are broken. It may also be
noted that such global symmetries have played a role in explaining
neutrino masses in SUSY models, discussed earlier \cite{ahmsw,by}.

We describe our model in detail and essential features of it
in Section \ref{sec:Mod}. We obtain the low-energy scalar potential
of our model in Section \ref{sec:sca} and argue that the sneutrinos
cannot acquire non-zero vevs. Since we need a non-zero scalar vev for
the field $X$ which is a carrier of lepton number, there are $\Delta L = 1$
interaction terms in our model. This and some other consequences of our
model are given in Section \ref{sec:con}. Section \ref{sec:fin} contains
conclusions.

\section{Essential features of the model}
\label{sec:Mod}

As it is explained in the introduction, we assume the conservation
of lepton number and R-symmetry at the Planck scale. We construct a Lagrangian
which is invariant under these symmetries.
Apart from the gauge kinetic terms, the Lagrangian of the model is
\begin{equation}
{\cal L} = \int Kd^4\theta + (\int Wd^2\theta + {\rm h.c.}),
	\label{eq:lag}
\end{equation}

\noindent
where $W$ and $K$ are the superpotential and the Kahler potential, 
respectively.

It is well known that in the low energy limit of a spontaneously
broken supergravity model, the cancellation of large contributions to 
the cosmological constant requires
the presence of at least one scalar field
(usually a singlet under the observable 
sector gauge group) with a vev of the order of Planck scale 
\cite{susy,nilles,sugra1}. 
If the  superpotentials for this (hidden sector) field and 
the observable sector fields are additive, 
then the cosmological constant is determined by the vevs of the
hidden sector field(s) \cite{sugra1}. Given such vevs, and with
appropriate choice of parameters in the hidden sector superpotential 
in such a scenario, supersymmetry can be broken at an intermediate scale with 
gravitino mass as low as  $\sim$ TeV. This in turn results in the
generation of SUSY-breaking soft terms of the order of gravitino mass
in the observable sector \cite{sugra2}. In our case, the field $S$ plays
this role in the hidden sector. Although the problem of cosmological constant
is not the main focus of this paper, we need the scalar vev of $S$ at the
Planck scale to attain the gravitino and scalar masses at the TeV scale.
We briefly comment on the cosmological constant in Section \ref{sec:con}.
As we shall see and it is already mentioned
in the introduction, the other hidden sector field, namely, $X$, 
on the other hand, has its vev at an intermediate
scale, and being a carrier of lepton number, has an altogether 
different role to play in the observable sector phenomenology.
Finally, the field $S^\prime$ should acquire zero vev in order for
the gaugino fields to acquire masses at the TeV scale.
We present below the explicit forms of the superpotential
and the Kahler potential, through which we can attain
consistent SUSY breaking.

The forms of the superpotential and the Kahler potential
follow from the conservation of R-symmetry and lepton
number, which we have assumed to be valid at the Planck scale.
The R-charges of the fields are assigned as:
$R(S) = 2$, $R(S^\prime) = 0$, $R(X) = 1$, $R(Q_i) = R(L_i) = R(U^c_i) =
R(N^c_i) = 0$, $R(D^c_i) = R(E^c_i) = 2$, $R(H_1) = 0$, $R(H_2) = 2$.
With this, the superpotential has the form
\begin{equation}
W = W_h + W_{\rm MSSM} +
    \frac{XX}{2M_P}N^c_ia_{ij}N^c_j,
	\label{eq:sp}
\end{equation}
where $M_P$ is the Planck scale and $a_{ij}$ are ${\cal O}$(1) constants,
with $a_{ij}=a_{ji}$. $W_{\rm MSSM}$ is defined in equation (\ref{eq:mssm}).
$W_h$ is the hidden sector part of the
superpotential which takes the following leading terms.
\begin{equation}
W_h = \Lambda^2S + \Lambda^\prime S^\prime S + qS^{\prime^2}S,
	\label{eq:hid}
\end{equation}
where $\Lambda\sim 10^{10-11}$ GeV, $\Lambda^\prime\sim$ TeV and
$q$ is a ${\cal O}$(1) constant. For phenomenological consistency of
our model we need the $\Lambda^\prime$ to be at the TeV scale. But in
this work we are not justifying it. There is another TeV scale parameter
in the superpotential, which is the $\mu$-term of $W_{\rm MSSM}$.
Here we are not addressing the origin of this term.
In the past there were attempts to understand why the $\mu$-parameter is
at the TeV scale \cite{ahmsw,gm}. We may follow these approaches and
address the origins of $\Lambda^\prime$ and $\mu$ terms, but that is
not main focus of this paper.
The last term of equation (\ref{eq:sp}), which arises from the conservation of
lepton number, is
especially noteworthy; such a non-renormalizable term can obviously
lead to $\Delta L =2$ neutrino masses once the scalar component of X 
acquires vev. The role of R-symmetry
in the superpotential is that it forbids the following lepton number
conserving terms,
such as: $\frac{S}{M_P}QU^cH_2,~\frac{S}{M_P}QD^cH_1,~\frac{S}{M_P}LE^cH_1,
~\frac{S}{M_P}LN^cH_2$. Since, as explained before, S acquires
a scalar vev of the order of Planck scale, these terms
generate unacceptably high masses for quarks and leptons, expect for
the top quark. However, terms of the form: $\frac{S^\prime}{M_P}QU^cH_2,~
\frac{S^\prime}{M_P}LE^cH_1$, etc are allowed both by lepton number and
R-symmetry conservations, but since $S^\prime$ acquires zero vev they
do not contribute to masses of fermions.
R-symmetry also forbids the term $SH_1H_2$, which
generates a high value for the $\mu$ parameter.
R-symmetry also forbids the term $XN^c$, which generates $\Delta L =1$
term in the low-energy regime through the scalar vev of $X$.
The role of lepton number conservation in the superpotential
is that it forbids the following
terms: $LLE^c$, $QLD^c$ and $LH_2$, which are R-symmetric invariant,
and these are the terms that should be avoided for the stability of
the LSP. It also forbids terms
such as $XX$ or $XXN^c$, which are in principle allowed by
R-symmetry.
Terms in higher powers of
$S$ as well as $X$ in $W$ are also absent via R-symmetry as well as
the assumption of lepton number conservation at the Planck scale.

Next, we suggest a specific form of the Kahler potential.
In general, the Kahler manifold is 
a real function of the fields $Y$ and  $Y^{\dagger}$, where
$Y=S,~S^\prime,~X$ in our case.
Since $X$ carries lepton number, here one has a Kahler potential where
$X$ enters only in the 
form $X^{\dagger}X$ (for the conservation of lepton number) and consider
a Kahler potential of the form
%
\begin{equation}
K = K_0(S,S^\prime,XX^\dagger) +
\sum_i \Phi^{\dagger}_i\Phi_i .
	\label{eq:kp}
\end{equation}
Here, $\Phi_i = Q_i,U^c_i,D^c_i,L_i,E^c_i,N^c_i,H_1,H_2$. $K_0$ is some
function of hidden sector fields, which
should be chosen so that the Kahler potential is invariant under
both the R-symmetry and lepton number. In general, this function
depends non-trivially on the hidden sector fields, and thus
the Kahler potential has non-minimal character \cite{sugra3}.
The Kahler potential can contain terms involving both hidden and
observable fields, such as:
\begin{equation}
\frac{S^\dagger S}{M_P^2}\Phi^{\dagger}_i\Phi_i, \quad
\frac{S^\prime}{M_P}\Phi^{\dagger}_i\Phi_i, \quad
\frac{X^\dagger X}{M_P^2}\Phi^{\dagger}_i\Phi_i.
	\label{eq:int}
\end{equation}
Among the above terms, the first two at most contribute to the SUSY-breaking
soft terms for the scalar fields and they do not generate undesired
$\Delta L = 1$ terms. So we do not consider them in the Kahler potential
as they do not effect the main conclusions of our work.
Moreover, the soft terms for scalar fields can be obtained from the
Kahler potential that we have chosen in equation (\ref{eq:kp}), and
it will be shown in Section \ref{sec:sca}. However, the last term of
equation (\ref{eq:int}) can give rise to $\Delta L = 1$ terms through
the scalar vev of $X$. But this term is suppressed by two powers of
Planck mass. In Section \ref{sec:con} we argue that in our model
$\Delta L = 1$ terms dominantly come due to the last term of
equation(\ref{eq:sp}). It is clear that this term in the superpotential
is suppressed by
one power of Planck mass, so it gives dominant effects compared
to the last one of equation (\ref{eq:int}). To present our ideas
in a simple fashion we omit this possible term in the Kahler potential.
The Kahler potential can also contain the term
$\frac{S^\dagger}{M_P}H_1H_2$ which is consistent with the lepton number
and R-symmetry conservations. This term effectively generates $\mu$-term
if the auxiliary vev of $S$ is non-zero. It will be shown below that the
auxiliary vev of $S$ is at the intermediate scale and so the effective
$\mu$ parameter is at the TeV scale. Such a $\mu$-term is already
there in the superpotential and here our main motivation is not on the
explanation of the origin of $\mu$-term. So without loss of generality the term
$\frac{S^\dagger}{M_P}H_1H_2$ can be excluded from the Kahler potential.
Notice that
the assumption of conservation of R-symmetry at the Planck scale,
enables us to avoid the
terms of the form $\frac{X^\dagger}{M_P^2}LLE^c$,
$\frac{X^\dagger}{M_P^2}QLD^c$ and $\frac{X^\dagger}{M_P}LH_2$ which,
via a vev of $X$, violate lepton number by one unit.

To completely specify our model, we need to fix the
gauge kinetic function which determines the interactions of gauge and gaugino
fields. We do not study them in detail since they do not effect our
conclusions. But to be phenomenologically consistent, gaugino fields
should have masses which are determined by the form of gauge kinetic function.
In our model its form is
\begin{equation}
F_{ab} = \delta_{ab}\left(\frac{1}{g_a^2} + \frac{1}{M_P}f_aS^\prime
+ \cdots\right),
	\label{eq:gkf}
\end{equation}
where $g_a$ are the three gauge couplings of the standard model gauge group,
$f_a$ are ${\cal O}(1)$ constants and the indices $a,b$ run over 1,2,3.
The dots in equation (\ref{eq:gkf}) are higher order terms which can be
neglected.
The second term of equation (\ref{eq:gkf}) gives masses to gauginos.
We will see below that the auxiliary vev of $S^\prime$ is at the
intermediate scale and hence the gaugino fields have masses of the
order of $\frac{\langle F_{S^\prime}\rangle}{M_P}\sim$ TeV.

So far, everything included in $W$ as well $K$ conserve
lepton number. Now, the very form of $W$ tells us that 
$\langle F_S\rangle =\langle\frac{\partial W}{\partial S}\rangle$ is on
the order of $\Lambda^2$ and $\langle F_{S^\prime}\rangle = \langle
\frac{\partial W}{\partial S^\prime}\rangle = \Lambda^\prime\langle S\rangle
+ 2q\langle S^\prime\rangle\langle S\rangle$ is at the intermediate
scale if $\langle S\rangle\sim M_P$ and $\langle S^\prime\rangle = 0$.
We will make the scalar vev of $S^\prime$ to be zero in order to get its
auxiliary vev at the intermediate scale, and this we require to make sure
that the gauginos have masses at the TeV scale which is explained in the
previous paragraph.
While $\langle F_X\rangle = \langle\frac{\partial W}{\partial X}\rangle =0$
if the right-handed sneutrinos $\tilde{N_i}$ have no vev,
something that we need to establish in order to eliminate the possibility of
$\Delta L = 1$ terms.

\section{Scalar potential}
\label{sec:sca}

Let us now consider the scalar potential of this theory and place
our claims about the vevs of $S$, $S^\prime$ and $X$ on firmer ground.
The reasons for $\langle S^\prime\rangle = 0$ and $\langle X\rangle\sim
\Lambda$ have already been explained. We will show below that in order
for the gravitino mass to be at the TeV scale we need $\langle S\rangle\sim
M_P$. These are the demands that we are making on the vevs of hidden sector
fields to be in consistent with the phenomenological masses of the
supersymmetric fields. One crucial thing to be shown is that the
sneutrino fields should not acquire non-zero vevs.
We show below that these demands can be satisfied by minimizing
the scalar potential and choosing appropriate values of the parameters
of the model. 
The contribution to the scalar potential from the
superpotential and the Kahler potential is given by \cite{susy}
\begin{equation}
V = M_P^4e^G[M_P^2 G_M K^{M\bar{N}}G_{\bar{N}} - 3],
        \label{eq:pot}
\end{equation}
where
\begin{equation}
G = \frac{K}{M_P^2} + \ln\left|\frac{W}{M_P^3}\right|^2,
	\label{eq:G}
\end{equation}
\noindent
$G_M = \frac{\partial G}{\partial\phi_M}$ and $G_{\bar{N}} =
\frac{\partial G}{\partial\phi^*_N}$, $\phi$ being a chiral superfield.
The matrix $K^{M\bar{N}}$ is the inverse of
$\frac{\partial^2K}{\partial\phi^*_N\partial\phi_M}$.

The vevs that we get after minimizing the scalar potential determine
the SUSY breaking of the theory.
SUSY breaking requires that, expressed in terms of these, the vev of
\begin{eqnarray}
{\cal F}_\phi = \frac{\partial W}{\partial\phi} + \frac{W}{M_P^2}
\frac{\partial K}{\partial\phi} 
\end{eqnarray}
\noindent 
should be non-zero for some hidden sector field(s) $\phi$ \cite{sugra4}. 
In our case, we have
$\langle {\cal F}_S\rangle = \Lambda^2+\frac{\Lambda^2\langle S\rangle}
{M_P^2}\langle\frac{\partial K}{\partial S}\rangle$, after putting
$\langle S^\prime \rangle =\langle \tilde{N}^c_i \rangle = 0$.
If one has $\langle S \rangle$ at the Planck scale, together with $\Lambda$ 
at an intermediate scale, one can not only have a non-zero
$\langle {\cal F}_S \rangle$ but also
ensure $\langle {\cal F}_S \rangle$ of an order 
which is required by a phenomenologically consistent 
SUSY spectrum. One of the consequences of SUSY breaking in supergravity is that
the gravitino field acquires a non-zero mass, which is given below
\begin{eqnarray}
m_{3/2}^2 &\sim & M_P^2e^{\langle G\rangle} =
e^{\frac{\langle K\rangle}
{M_P^2}}\frac{\langle W\rangle^2}{M_P^4} = e^{\frac{\langle K\rangle}{M_P^2}}
\frac{(\Lambda^2\langle S\rangle)^2}{M_P^4}.
\end{eqnarray}
\noindent
The mass of gravitino can be of the order of TeV provided if
$\langle S\rangle\sim M_P$. 

Substituting the forms of $K$ and $W$ in equations (\ref{eq:pot}) and
(\ref{eq:G}), one obtains the form of the scalar potential \cite{sw} as
\begin{equation}
V = V_0 + V_1
\end{equation}
with
\begin{eqnarray}
V_{0} &=& e^{K/M_P^2}\left\{
K_0^{S\bar{S}}\left|\Lambda^2+\Lambda^\prime S^\prime +qS^{\prime^2}\right|^2
+ K_0^{S\bar{S}}\frac{\partial K_0}{\partial S}\left(\frac{\partial K_0}
  {\partial S}\right)^* \frac{W_h W_h^*}{M_P^4}
+ K_0^{S^\prime \bar{S^\prime}}\left|\Lambda^\prime S+2qS^\prime S \right|^2 -
  \frac{3}{M_P^2}W_h W_h^*
	\right. \nonumber \\
&& \left.
+ K_0^{S^\prime \bar{S^\prime}}\frac{\partial K_0}{\partial S^\prime}
  \left(\frac{\partial K_0}{\partial S^\prime}\right)^* \frac{W_h W_h^*}{M_P^4}
+ \left(K_0^{S\bar{S}}(\Lambda^2+\Lambda^\prime S^\prime +qS^{\prime^2})^*
     \frac{\partial K_0}{\partial S}\frac{W_h}{M_P^2}
+ K_0^{S \bar{S^\prime}}(\Lambda^2+\Lambda^\prime S^\prime +qS^{\prime^2})
  (\Lambda^\prime S+2qS^\prime S)^*
	\right. \right. \nonumber \\
&& \left. \left.
+ K_0^{S \bar{S^\prime}}(\Lambda^\prime S+2qS^\prime S)^*
  \frac{\partial K_0}{\partial S}\frac{W_h}{M_P^2}
+ K_0^{S\bar{S^\prime}} (\Lambda^2+\Lambda^\prime S^\prime +
  qS^{\prime^2})\left(\frac{\partial K_0}{\partial S^\prime}\right)^*
  \frac{W_h^*}{M_P^2}
+ K_0^{S\bar{S^\prime}}\frac{\partial K_0}{\partial S}
  \left(\frac{\partial K_0}{\partial S^\prime}\right)^* \frac{W_h W_h^*}{M_P^4}
	\right. \right. \nonumber \\
&& \left. \left.
+ K_0^{S^\prime \bar{S^\prime}}(\Lambda^\prime S+2qS^\prime S)
  \left(\frac{\partial K_0}{\partial S^\prime}\right)^* \frac{W_h^*}{M_P^2}
+ {\rm h.c.}\right)
	\right. \nonumber \\
&& \left.
+ K_0^{X\bar{X}}\frac{\partial K_0}{\partial X}\left(\frac{\partial K_0}
  {\partial X}\right)^* \frac{W_h W_h^*}{M_P^4}
+ \left( K_0^{S\bar{X}}(\Lambda^2+\Lambda^\prime S^\prime +qS^{\prime^2})
  \left(\frac{\partial K_0}{\partial X}\right)^* \frac{W_h^*}{M_P^2}
+ K_0^{S\bar{X}}\frac{\partial K_0}{\partial S}\left(\frac{\partial K_0}
  {\partial X}\right)^* \frac{W_h W_h^*}{M_P^4}
	\right. \right. \nonumber \\
&& \left. \left.
+ K_0^{S^\prime \bar{X}}(\Lambda^\prime S+2qS^\prime S)
  \left(\frac{\partial K_0}{\partial X}\right)^* \frac{W_h^*}{M_P^2}
+ K_0^{S^\prime \bar{X}}\frac{\partial K_0}{\partial S^\prime}
  \left(\frac{\partial K_0}{\partial X}\right)^* \frac{W_h W_h^*}{M_P^4}
+ {\rm h.c.} \right)
\right\}
\end{eqnarray}
and
\begin{eqnarray}
V_1 &=& e^{K/M_P^2}\left\{\left(\frac{\partial W_{\rm MSSM}}{\partial\Phi_i}
  \right)^* \frac{\partial W_{\rm MSSM}}{\partial\Phi_i}
+ m_0^2(S,S^\prime)\Phi^*_i\Phi_i
+ \left(A_1(S,S^\prime)\frac{\partial W_{\rm MSSM}}{\partial\Phi_i}\Phi_i
+ A_2(S,S^\prime)W_{\rm MSSM}
	\right. \right. \nonumber \\
&& \left. \left.
+ B_N(S,S^\prime,X)\tilde{N}^c_ia_{ij}\tilde{N}^c_j
+ \frac{XX}{M_P}\left(\frac{\partial W_{\rm MSSM}}{\partial\tilde{N}^c_i}
  \right)^* a_{ij}\tilde{N}^{c}_j
+ \frac{XXX^*X^*}{2M^2_P}a_{ij}a^*_{ik}\tilde{N}^c_j\tilde{N}^{c^*}_k
+ {\rm h.c.}\right)
\right\},
	\label{eq:v1}
\end{eqnarray}
where
\begin{eqnarray}
m_0^2(S,S^\prime) &=& \frac{W_h W_h^*}{M_P^4}, \quad
A_1(S,S^\prime) = \frac{W_h^*}{M_P^2},
	\nonumber \\
A_2(S,S^\prime) &=& K_0^{S\bar{S}}\frac{(\Lambda^2+\Lambda^\prime S^\prime +
  qS^{\prime^2})^*}{M_P^2}\frac{\partial K_0}{\partial S}
  + K_0^{S\bar{S}}\frac{\partial K_0}{\partial S}\left(\frac{\partial K_0}
  {\partial S^\prime}\right)^* \frac{W_h^*}{M_P^4}
  + K_0^{S\bar{S^\prime}}\frac{(\Lambda^\prime S+2qS^\prime S)^*}{M_P^2}
  \frac{\partial K_0}{\partial S}
	\nonumber \\
&&
  + K_0^{S^\prime\bar{S}}\frac{(\Lambda^\prime S+2qS^\prime S)^*}{M_P^2}
  \frac{\partial K_0}{\partial S^\prime}
  + K_0^{S\bar{S^\prime}}\frac{W_h^*}{M_P^4}\frac{\partial K_0}{\partial S}
  \left(\frac{\partial K_0}{\partial S^\prime}\right)^*
  + K_0^{S^\prime\bar{S^\prime}}\frac{(\Lambda^\prime S+2qS^\prime S)^*}{M_P^2}
  \frac{\partial K_0}{\partial S^\prime}
	\nonumber \\
&&
  + K_0^{S^\prime\bar{S^\prime}}\frac{W_h^*}{M_P^4}\frac{\partial K_0}
  {\partial S^\prime}\left(\frac{\partial K_0}{\partial S^\prime}\right)^*
  - 3\frac{W_h^*}{M_P^2},
	\nonumber \\
B_N(S,S^\prime,X) &=& \frac{XX}{2M_P}A_2(S,S^\prime)
  + \frac{X}{M_P}K_0^{X\bar{S}}(\Lambda^2+\Lambda^\prime S^\prime +
  qS^{\prime^2})^*
  + \frac{X}{M_P}K_0^{X\bar{S}}\left(\frac{\partial K_0}{\partial S}\right)^*
  \frac{W_h^*}{M_P^2}
	\nonumber \\
&&
  + \frac{X}{M_P}K_0^{X\bar{S^\prime}}(\Lambda^\prime S+2qS^\prime S)^*
  + \frac{X}{M_P}K_0^{X\bar{X}}\left(\frac{\partial K_0}{\partial X}\right)^*
  \frac{W_h^*}{M_P^2}
  + \frac{XX}{M_P}\frac{W_h^*}{M_P^2}.
	\label{eq:par}
\end{eqnarray}
It should be noted that the chiral superfields and their scalar
components have been denoted by the same set of symbols here.
The contribution to scalar potential from the gauge kinetic function
is neglected here, since the scalar vev of $S^\prime$ is made to be
zero which is shown below.
By minimizing the scalar potential we have to achieve the vevs of hidden
sector fields as: $\langle S\rangle\sim M_P$, $\langle X\rangle\sim \Lambda$
and $\langle S^\prime\rangle = 0$. We can achieve this by choosing an
appropriate form of the function $K_0(S,S^\prime, X)$. There are six
independent double derivatives and three independent single derivatives of
$K_0$ in the scalar potential. We can fix them in such a way
that the desired vevs of hidden sector fields arise after minimizing
the potential. In deriving $V_1$, we have used $\langle \Phi_i\rangle
\ll \langle X\rangle\ll \langle S\rangle$, and terms suppressed by higher
powers of $M_P$ have been neglected. Substituting the vevs of the hidden
sector fields in $V_1$, we get the low-energy scalar potential. The first
term in equation (\ref{eq:v1}) is the F-term contribution to the scalar
potential of MSSM. Remaining terms in equation (\ref{eq:v1}) are SUSY-breaking
soft masses. The soft masses, which are given in equation (\ref{eq:par}),
are determined by the high scale parameters of our model. It can be noticed
that these soft masses are at the TeV scale after plugging the vevs
of hidden sector fields in their respective formulas. So our proposed
model is consistent with SUSY breaking as it produces a viable low-energy
scalar potential. For simplicity, we list the vevs of the scalar and
auxiliary components of the additonal superfields that we require
in our model, in Table \ref{T:t1}. In Table \ref{T:t2}
we present a list of sources for soft terms in the SUSY Lagrangian,
completely determined by the superpotential and
scalar potential, and a clear demonstration of how they are governed 
by the vevs of the hidden sector fields.

While our main conclusions
depend critically on various vevs in the hidden sector, it may be
argued that their values can be altered through gravitational effects.
However, these effects should in general be suppressed by ${\cal O}
(\Lambda /M_P)$ and therefore can be ignored in the preliminary proposal.
%

\begin{table}
\begin{center}
\begin{tabular}{||c|c|c||} \hline
Field & Scalar vev & Auxiliary vev \\ \hline
$S$   & $\sim M_P$ & $\sim\Lambda^2$ \\
$S^\prime$ & 0     & $\sim\Lambda^2$ \\
$X$   & $\sim\Lambda$ & 0 \\
$N^c$ & 0 & 0 \\ \hline

\end{tabular}
\end{center}
\caption{vev's of the scalar and auxiliary components of the addtional
chiral superfields.}
	\label{T:t1}
\end{table}
\begin{table}
\begin{center}
\begin{tabular}{||c|c|c||} \hline

Parameter & Source & Order of \\
          &        &  magnitude \\ \hline
$m^2_0$   & $m^2_0(S,S^\prime)$ in $V_1$ & TeV$^2$ \\
$A$       & $A_1(S,S^\prime),~A_2(S,S^\prime)$ in $V_1$ & TeV \\
$B_\mu$   & $\mu A_{1(2)}(S,S^\prime)$ in $V_1$ & TeV$^2$ \\
$m_{1/2}$ & $\frac{F_{S^\prime}}{M_P}$ from gauge kinetic terms & TeV \\
\hline

\end{tabular}
\end{center}
\caption{The different parameters of low energy SUSY and their sources.}
	\label{T:t2}
\end{table}

As mentioned above, we need to ascertain
that neither the left- nor the right-chiral sneutrinos develop 
any vev. To ensure this, one has to fulfill the minimization conditions
\cite{rm} for the low-energy scalar potential $V_1$
(i.e. vanishing of the first derivatives,
positivity of the eigenvalues of the second derivatives, etc.) for
$\langle {\tilde \nu}_i\rangle$ and  $\langle {\tilde N}_i\rangle$
simultaneously. We have checked that such solutions can be
guaranteed for appropriate values of 
the parameters in $A_{1(2)}$, $B_N$, $B_{\mu}$ and
$a_{ij}$ as well as the vev of $X$ and $S$ . Ensuring this is relatively
easy, since the right-chiral sneutrinos do not occur in quartic 
terms (except those suppressed by $M^4_P$), and can develop vev only 
through terms linear in the vevs of the left-chiral sneutrinos.  
Thus it is enough to make the latter zero through an appropriate choice
of parameters.

\section{Consequences}
\label{sec:con}

One of the consequences of our model is that we get $\Delta L = 2$
mass terms for neutrinos. After giving vev to $X$ the last term of
equation (\ref{eq:sp}) gives right-handed Majorana neutrino
mass of the form $M_R \sim \frac{\langle X\rangle^2}{M_P}\sim$ TeV.
If the neutrino Yukawa couplings: $Y_\nu \sim 10^{-7}$, then the Dirac
mass for neutrinos turns out to be $m_D \sim Y_\nu v_2 \sim 10^{-4}$ GeV,
where $\langle H_2^0\rangle = v_2$. It may be legitimate to take the
neutrino Yukawa couplings of ${\cal O}(10^{-7})$, since this is the same
as that of the electron Yukawa coupling and we do not understand why
the electron mass is that much small. If we put the Yukawa couplings
of electron and neutrinos on same footing, we can explain the smallness
of neutrino masses. In our model, since the right-handed neutrino mass
is much heavier than the Dirac mass for neutrinos, the seesaw mass formula
for light neutrinos is $m_\nu = -m_D^2/M_R \sim$ 0.1 eV. This is the
right magnitude for the neutrino mass which has been estimated from the
neutrino oscillation
experiments. So we can explain consistently the Majorana nature of neutrino
and its smallness of mass in our model.

Another consequence of our model is that the fermionic state belonging
to the chiral field $X$, which we denote as $\psi_X$, becomes massless
at the tree level. This statement follows from the last term
of equation (\ref{eq:sp}), since the right-chiral sneutrinos are made
to have zero vevs. Our model requires a non-zero vev for $X$ and the last term
of equation (\ref{eq:sp}) generates an effective term of the form
\begin{equation}
\frac{\langle X\rangle}{M_P}XN^cN^c
	\label{eq:L=1}
\end{equation}
in the superpotential. Through this $\Delta L = 1$ term and the neutrino
Yukawa couplings, $\psi_X$ mixes with the neutralino
states through  one-loop diagrams, if the neutral $H_2^0$
state acquires vev. This loop diagram gives very small contribution and
the mass eigenvalue of $\psi_X$ is less than the mass of any supersymmetric
particle. As a result of
this, through the same one-loop diagram the lightest neutralino of MSSM
can decay to $\psi_X$ and a neutral Higgs boson. We have found that
the decay width of this process is approximately given by
\begin{equation}
\Gamma \sim \frac{1}{8\pi}\left(\frac{g}{16\pi^2}Y_\nu^2\frac{\langle X\rangle}
{M_P}\right)^2 m_{\chi^0},
\end{equation}
where $g$ is the SU(2) gauge coupling strength and $m_{\chi^0}$ is
the mass of lightest neutralino. For typical values of parameters
in the above equation, the lifetime of lightest neutralino is
$\tau = \frac{1}{\Gamma} \sim 2\times 10^{11}$ years. This value
is one order of magnitude greater than the age of the universe.
Such a long lifetime of neutralino in our model is due to the
fact that the effective $\Delta L = 1$ term in the superpotential,
equation (\ref{eq:L=1}), is suppressed by a factor of
$\frac{\langle X\rangle}{M_P}\sim 10^{-7}$ and the neutrino Yukawa
couplings give further suppression in the loop induced decay.

In addition,
the last term of equation (\ref{eq:sp}) generates some scalar interactions
in the low-energy scalar potential, which are the last three terms of
equation (\ref{eq:v1}). They can generate $\Delta L = 1$ terms through
the vev of $X$, which have the following schematic forms:
\begin{equation}
\frac{\langle X\rangle}{M_P}AX\tilde{N}^c\tilde{N}^c, \quad
\frac{\langle X\rangle}{M_P}\frac{(XX)^*}{M_P}X\tilde{N}^c\tilde{N}^{c^*},\quad
\frac{\langle X\rangle}{M_P}X(LH_2)^*\tilde{N}^c.
\end{equation}
Here, $A\sim$ TeV. All the above three terms have
a suppression factor of $\frac{\langle X\rangle}{M_P}\sim 10^{-7}$.
We have found that they do not induce any tree level decay of the
neutralino state. The loop induced decays due to them will have an
additional suppression of neutrino Yukawa couplings. Thus, in our model
the last term of equation (\ref{eq:sp}) is only the source for $\Delta L = 1$
terms. All of them are suppressed by Planck mass and the
lepton number violation by one unit is confined only in the neutrino
sector. Because of these reasons, the lightest neutralino of MSSM can
have lifetime exceeding the age of the universe, and it can
still provide a candidate for the dark matter content. Now it can be
easily justified why the last one of equation (\ref{eq:int}) is
neglected in the Kahler potential: since that term is suppressed by two
powers of Planck mass, it gives sub-leading contributions to the
$\Delta L = 1$ terms and to the neutralino decay.

The low-energy scalar potential is also a consequence of our model.
The SUSY-breaking soft parameters in the scalar sector come out
in the phenomenologically expected range, as
listed in Table \ref{T:t2}. An intermediate scale vev of the auxiliary
component of $S^{\prime}$ justifies gaugino masses in the same
scale as well, through $S^{\prime}$ participating in the gauge kinetic
function. The parameter $B_\mu$, which is the coefficient of 
the bilinear term $H_1H_2$
in the scalar potential, is at the TeV scale provided the $\mu$ parameter
lies around the TeV scale. While we have not justified the value of $\mu$ 
in this work, an explanation of the twin parameters $\mu$ and
$\Lambda^{\prime}$ in the proposed scenario, both belonging to the
superpotential and still around 
the electroweak scale, is hoped to come from a deeper understanding
of the `$\mu$-problem'.

It may be re-tread that the suggested orders of magnitude of the scalar 
and auxiliary components vevs of the fields $X$, $S$ and $S^{\prime}$ are 
all consistent with observable sector SUSY breaking parameters, being all
in the TeV range. It is intimately related to the fact that the same set of 
choices yields a gravitino mass on the same order. 

Another interesting possibility which is kept alive by such choice concerns
the cosmological constant which, as is well-known, needs to be fine-tuned to 
a miniscule value in SUSY scenarios. The dominant contribution 
to this constant in our model comes from the part $V_0$ 
of the scalar potential.
After giving vevs to hidden sector fields, the first four
lines of $V_0$ give a contribution of the order of $\Lambda^4$ and the
last two lines give contribution of the order of $\Lambda^2{\rm TeV^2}$.
Although we are not claiming to solve the cosmological constant problem, 
the choice of fields and orders of their vevs make it possible to 
envision the mutual cancellation of the dominant terms contributing to it,
by proper adjustment of the dimensionless parameters occurring in the
Kahler potential.

\section{Conclusions}
\label{sec:fin}

To conclude, we have suggested a supersymmetric scenario where a field 
carrying lepton number but otherwise immune to standard model interactions
can generate $\Delta L = 2$ neutrino mass terms and also keeps the
lightest neutralino of MSSM long lived particle. In this scenario we have
proposed three kinds of hidden sector fields: $S$, $S^\prime$ and $X$, where
only $X$ carries a lepton number.
The purposes served by  these fields are (i) the generation of 
$\Delta L = 2$ mass terms for neutrinos, (ii) the elongation of the
the lifetime of the LSP, decaying through $\Delta L = 1$ interactions,
beyond the present age of the universe, and (iii) the
occurrence of  SUSY breaking parameters around the TeV scale, thus
yielding a phenomenologically viable SUSY spectrum. The lepton
number carried by $X$ as well as the R-charge assignments of various
fields ensure this, both the charges being broken at energies
below the Planck scale, along a line frequently taken in SUSY models
of neutrino masses. It is also demonstrated that the scenario proposed here
can accommodate a cancellation of the leading terms contributing to the value
of the cosmological constant. This shows the potency of supergravity
theories in reconciling seesaw masses for neutrinos with the
observed cold dark matter of the universe, and underscores the
importance of attempts to derive scenarios such as the aforesaid
one from more fundamental principles.

{\bf Acknowledgment:} This work was partially supported by the Department of 
Atomic Energy, Government of India, through a project granted in the Xth
Five-years Plan. SSG acknowledges the hospitality of  Harish-Chandra
Research Institute during the early stage of this work, while BM and RS
thank Indian Association for the Cultivation of Science, for 
hospitality at a later stage.


\begin{thebibliography}{10}

\bibitem{susy}
For general introductions, see, for example,
H.P. Nilles, Phys. Rep. {\bf 110}, 1 (1984);
{\it Perspectives in Supersymmetry}, G. Kane (Ed.),
World Scientific, Singapore (1998).

\bibitem{dm}
N. Arkani-Hamed, A. Delgado and G.F. Giudice, Nucl. Phys. {\bf B741},
108 (2006);
S.F. King and J.P. Roberts, [arXiv: hep-ph/0603095]; and references therein.

\bibitem{seesaw}
P. Minkowski, Phys. Lett. B {\bf 67}, 421 (1977);
T. Yanagida, in {\it Proceedings of the workshop on unified theory and
baryon number in the universe}, KEK, March 1979, eds. O. Sawada and
A. Sugamoto;
M. Gell-Mann, P. Ramond and R. Slansky, in {\it Supergravity}, Stonybrook,
1979, eds. D. Freedman and P. van Nieuwenhuizen;
S.L. Glashow, \emph{The future of elementary particle physics}, in
   \emph{Proceedings of the 1979 Carg{\`e}se Summer Institute on
Quarks and Leptons} (M.~L{\'e}vy et al. eds.), Plenum Press, New York,
1980, pp.~687;
R.N. Mohapatra and G. Senjanovi{\'c},
Phys.\ Rev.\ Lett.\  {\bf 44}, 912 (1980).

\bibitem{rpv}
H.E. Haber and G.L. Kane, Phys. Rep. {\bf 117}, 75 (1985).

\bibitem{gut}
See, for example,
R.N. Mohapatra, Phys. Rev. D {\bf 54}, 5728 (1996);
Z. Berezhiani and Z. Tavartkiladze, Phys. Lett. B {\bf 409}, 220 (1997);
H.S. Goh, R.N. Mohapatra and S. Ng, Phys. Rev. D {\bf 68}, 115008 (2003);
U. Sarkar, Phys. Lett. B {\bf 622}, 118 (2005).

\bibitem{Martin}
S.P. Martin, Phys. Rev. D {\bf 46}, 2769 (1992); {\it ibid.} {\bf 54},
2340 (1996).

\bibitem{sugra}
E. Cremmer {\it et al.}, Nucl. Phys. {\bf B212}, 43 (1983);
A. Chameseddine, R. Arnowitt and P. Nath, {\it Applied N = 1 Supergravity}
(World Scientific, Singapore, 1984).

\bibitem{ahmsw}
N. Arkani-Hamed, L. Hall, H. Murayama, D. Smith and N. Weiner, Phys. Rev.
D {\bf 64}, 115011 (2001); [arXiv: hep-ph/0007001]

\bibitem{by}
K.S. Babu and T. Yanagida, Phys. Lett. B {\bf 491}, 148 (2000).
F. Borzumati and Y. Nomura, Phys. Rev. D {\bf 64}, 053005 (2001);
F. Borzumati, K. Hamaguchi and T. Yanagida, Phys. Lett. B {\bf 497},
259 (2001);

\bibitem{ms}
B. Mukhopadhyaya, P. Roy and R. Srikanth, Phys. Rev. D {\bf 73},
035003 (2006);
B. Mukhopadhyaya and R. Srikanth, Phys. Rev. D {\bf 74}, 075001 (2006).

\bibitem{bb}
L. Bento and Z. Berezhiani, Phys. Rev. Lett. {\bf 87}, 231304 (2001).

\bibitem{nilles} H.P. Nilles, [arXiv: hep-ph/0603095].

\bibitem{sugra1} E. Cremmer, B. Julia, J. Scherk, S. Ferrara, L. Girardello
and P. van Nieuwenhuizen,
Nucl. Phys. {\bf B147}, 105 ( 1979);
E. Cremmer, S. Ferrara, C. Kounnas and D.V. Nanopoulos, Phys. Lett. B {\bf 133}, 61 (1983).

\bibitem{sugra2} L. Hall, J. Lykken and S. Weinberg, Phys. Rev. D {\bf 27}, 2359 (1983) ;
H.P. Nilles, Nucl. Phys. {\bf B217}, 366 (1983).

\bibitem{gm}
G.F. Giudice and A. Masiero, Phys. Lett. B {\bf 206}, 480 (1988).

\bibitem{sugra3} J. Bagger, Nucl. Phys. {\bf B211}, 302 (1983); 
J. Ellis, C. Kounnas and D.V. Nanopoulos, Nucl. Phys. {\bf B247}, 373 (1984).

\bibitem{sugra4} A. Brignole, L.E. Ibanez and C. Munoz, in {\it Prespectives in Supersymmetry},
G. Kane (Ed.), World Scientific, Singapore, 1998.

\bibitem{sw}
The method for doing this is exemplified in, for example,
S.K. Soni and H.A. Weldon, Phys. Lett. B {\bf 126}, 215 (1983);
see also \cite{sugra4}.

\bibitem{rm}
S. Roy and B. Mukhopadhyaya, Phys. Rev. D {\bf 55}, 7020 (1997).

\end{thebibliography}
\end{document}